\documentclass{PoS}

\title{Testing for a class of ULGRBs using Swift GRBs}

\ShortTitle{ULGRB class}

\author{\speaker{Michel Bo\"er}\\
        ARTEMIS (CNRS/UNS/OCA), Nice, France\\
        E-mail: \email{Michel.Boer@unice.fr}}
\author{Bruce Gendre\\
        ARTEMIS (CNRS/UNS/OCA), Nice\\
        E-mail: \email{Bruce.Gendre@gmail.com}}
\author{Giulia Stratta\\
        Universit\`a degli Studi di Urbino, Carlo Bo, Urbino, Italy\\
        E-mail: \email{giulia.stratta@uniurb.it}}

\abstract{The question of whether ultra-long GRBs form a population different from that of "regular" long GRBs has been much debated recently and during the conference. We discuss here the data
and the evidence that lead to the conclusion that indeed ultra-long GRBs form a different class of high energy transients. The sample of ultra-long GRBs is still poor and the discussion on their origin remain opens, though they might be the signature of PopIII stars. We urge that the design of new instrumentation, such as the SVOM satellite, takes into account the need for the detection of distant ultra-long GRBs.  \ }

\FullConference{Swift: 10 Years of Discovery,\\
		2-5 December 2014\\
		La Sapienza University, Rome, Italy }

\begin{document}

\section{Introduction}
The problem of the duration of the GRB prompt phase and its connection with their possible origin(s) is almost as old as their discovery. With the advent of Swift and the possibility to detect the prompt emission also in X-rays for some long GRBs, the burst durations extend over 6 orders of magnitude. This is only one of the multiple facet of their diversity as their variability extend over 6 decades and the energy over 8 decades if we consider the afterglow phase. Based on their duration they have been categorized in 2 classes \cite{kou93,dez92}, respectively the short and long GRBs (hereafter sGRBs and lGRBs). Though this classification is based on both their spectral and temporal properties, it uses only properties in the observer frame, being a problem for objects that have been observed from redshifts 0.1 to almost 9 (see also \cite{sie14, reg15} for an analysis of sGRBs in the rest frame). However, this categorization has proven to be useful, and lead to strong evidences for an association of sGRBs with compact binary coalescences \cite{eic89}, and lGRBs with the collapse of a very massive star to a black hole \cite{woo93}. This classification is also correlated with studies on the localization of the events with respect to the host galaxy ( e.g. \cite{fon13}).

The collapsar model can explain the amount of energy needed to supply a lGRB, and it is quite effective to explain many of the properties of these sources, such as the supernova associations \cite{hjo03, sta03} or the observation of stellar winds around the source \cite{gen04,gen05}. However, the lack of hydrogen in the envelope lead to the proposal that the progenitor of lGRBs should be a Wolf-Rayet star  (e.e. \cite{che99}).

This explanation is much more difficult with the so called ULGRBs (Ultra-Long Gamma-ray Bursts) as their duration is much larger than 1000s, and even above 20000s. To date we know 4 candidate ULGRBs, namely GRB 110209A \cite{gen13, str13, lev13}, GRB 101225A \cite{lev13, tho11}, GRB 121027A \cite{lev13, Wu13} and GRB 130925A \cite{pir14}.

The question of the origin of these events has been much debated during this conference and in the literature. Gendre et al. \cite{gen13} have proposed that a blue supergiant as the progenitor could provide the matter reservoir needed to feed sources of such extreme duration. Levan et al. \cite{lev13} reached the same conclusion, while Margutti et al. proposed a new class of soft and ultra-long events \cite{mar14}. Recently \cite{boe15} made a careful statistical analysis of set of long GRBs well observed over very long time and found that the probability that ULGRBs are an extension of the lGRB class is very low, therefore they conclude that they form a different population. However, Virgili et al. \cite{vir13} reached the opposite conclusion, and claimed that ULGRBs are the tail of the lGRB distribution rather than event with different progenitors.

One can also try to use other estimators of the GRB duration, such as modeling the X-ray light curve and using a theoretical model to estimate the duration of the central engine activity. The result of \cite{zha13} is that the activity of GRBs span the range between 0.1 and $10^6$s, rather the observed peak distribution around 30s observed using the usual $\rm{T}_{90}$ measure.

Another point to add to the analysis is that all the four ULGRBs observed to date display a thermal emission during the whole prompt phase, a very unusual feature in GRB spectra.

Whatever the case, ULGRBs being very long GRBs or a different classes, we are left with several questions. If ULGRBs are very long lGRBs, then the need arises to explain with the same mechanism bursts that last few seconds to more than 20000 s and with thermal emission in few (usually very long duration) cases. If ULGRBs form a different class, then a credible new progenitor, and/or physical mechanism has to be found to explain it.

Here we present the results of a statistical analysis aimed at determined whether or not ULGRBs, as a whole, form a different population or not. 

\begin{figure*}
\begin{center}
\includegraphics[width=8cm]{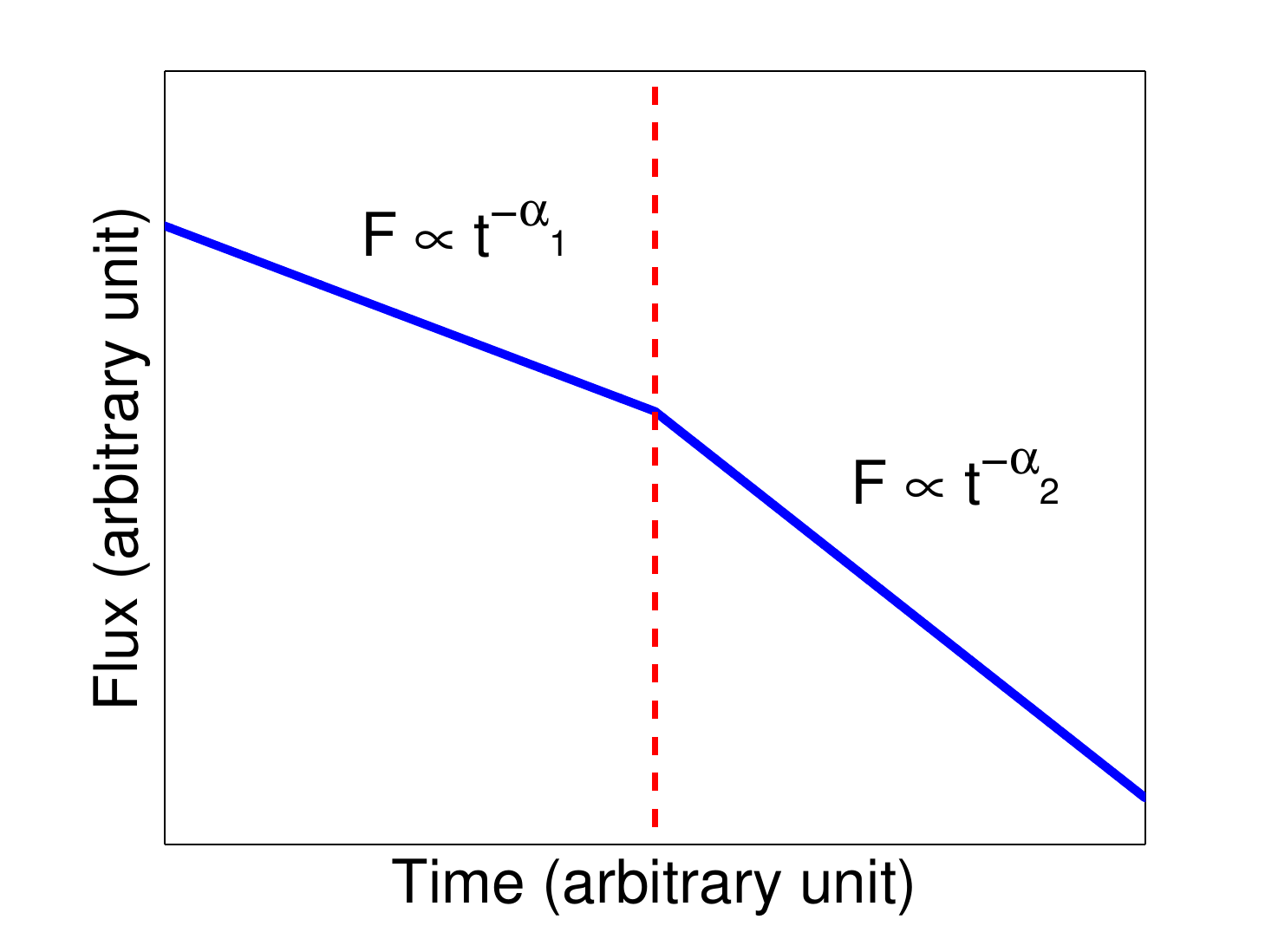} 
\includegraphics[width=8cm]{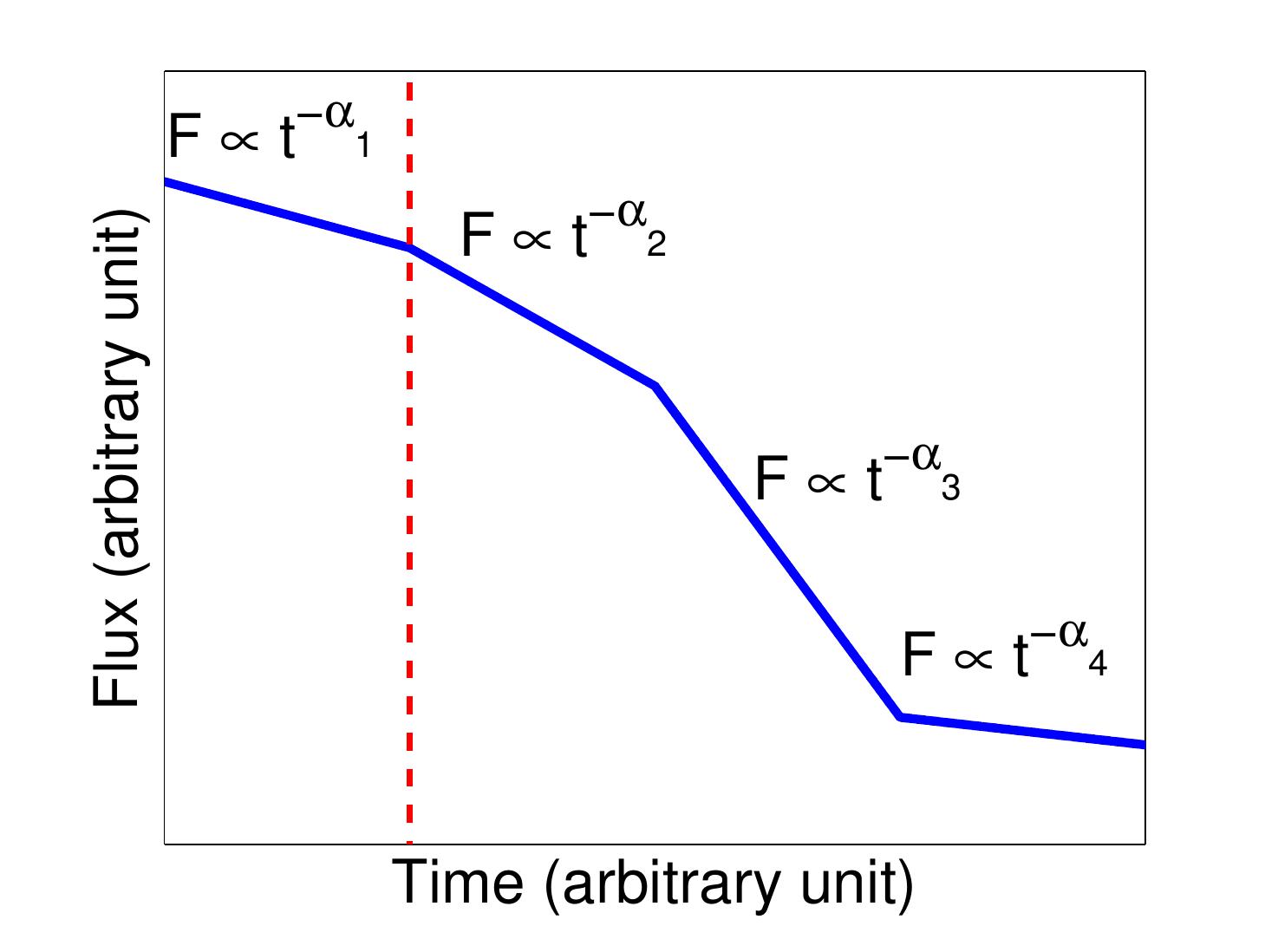}
\caption{Template light curve with the notation in use in \cite{boe15}. We present two examples. Above: one single break. Our method allow to discriminate if this break is prompt related or jet related. Below: a complex 3-break light curve. In this example, the start of the steep decay is the second segment if $\alpha _2 > 2.2$ (see more details in \cite{boe15}). 
\label{fig_template}}. 
\end{center}
\end{figure*}

\section{The new duration measure $T_X$}

Following the work done in \cite{boe15} we use the online Swift XRT GRB catalog \footnote{http://www.swift.ac.uk/xrt\_live\_cat/}\cite{eva09}. To model the overall shape of the light curves multiple power law segments $f(t)=kt^{-\alpha}$ are assumed. Flare episodes are considered as extra components in the XRT GRB catalog analysis, and removed during the estimation of the power law parameters. We have automated the analysis to extract the start time of the steep decay that we define to have $\alpha > 2.2$. The advantages of this method are 1) that it uses only data from a single instrument, therefore it is spectrally consistent, 2) that it enables us to study GRB of duration larger than 1000s over several orbits. We have removed GRBs for which the follow-up starts more than 500s after trigger time, and we do not consider flares or late-time activity. The method is illustrated in Fig. \ref{fig_template} and in \cite{boe15}.

We are left with a final sample of 207 events, out of the 243 taken from the catalog. 

We have a specific case, GRB 130925A that has a claimed T$_{90}$ duration of more than 20ksec, but our $T_X$ gives 149sec. This is probably due to a strong flaring activity. In any case we have considered for the analysis the two cases, with and without GRB 130925A.

The mean duration of $T_X$ = 137s and the median 119s. The minimum duration is 49s (but depends on the construction of the sample), while the maximum is 25400s. We have tested the different distributions:

\begin{itemize}
\item Log normal distribution
\item Log normal distribution excluding data above 300s
\item Generalized Extreme Value Distribution
\end{itemize}

\begin{figure*}
\begin{center}
\includegraphics[height = 8cm]{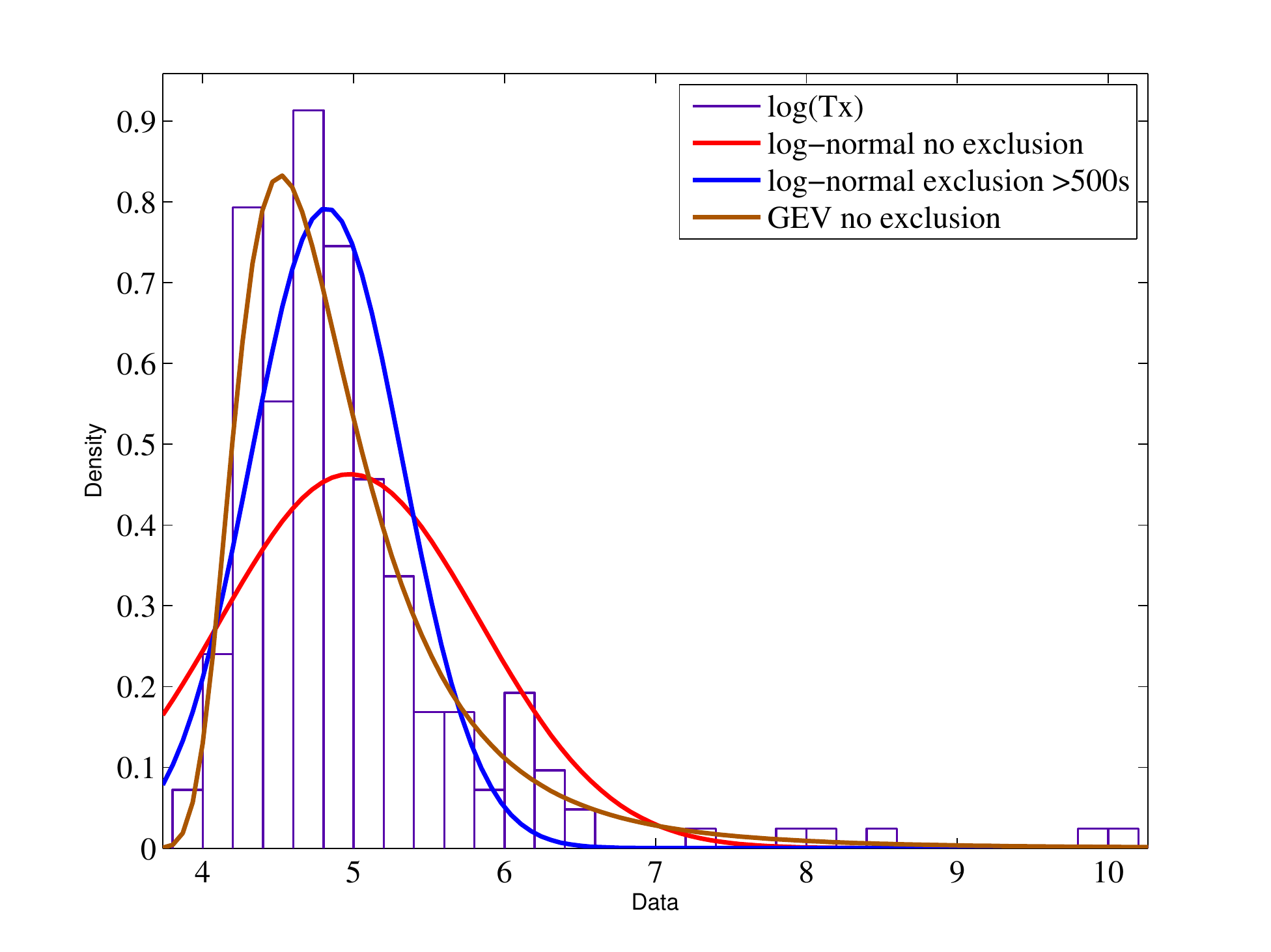} 
\caption{The probability density function of the logarithm of $T_X$ from \cite{boe15}. We have superimposed three different fits; red curve: log-normal model on the whole sample; blue curve: log-normal distribution, sample restricted to $T_X < 300$s; brown curve: generalized extreme value distribution, whole sample.\label{fig1}}
\end{center}
\end{figure*}

The distribution of the logarithm of $T_X$ as well as the above mentioned fits are plotted in Fig.\ref{fig1}.

The log-normal distribution on the whole set gives a very bad fit, and the null hypothesis is rejected to a high level (its probability is $2 \times 10^{-5}$). If we truncate the distribution and retain only the events above 300s, we do not get an acceptable fit either. Interestingly, it is by removing the events above 300s that we fit nicely the distribution. For example, keeping only events for which $T_x < 300$ s, we get $\chi^2=13$ for 15 d.o.f. (while for $T_x<2000$s we got $\chi^2=150$ for 14 d.o.f.). This could indicate that our limit for ULGRBs ($10^4$ s) is too conservative, and a value of $T_X > 10^3$ s should be more representative of this new class.

We get a better fit than the log-normal law with the GEV distribution. However it is still non ideal because of an excess of events with $T_X$ around 400-500s. The parameters are $\mu = 1.99 \pm 0.02$ (the location, i.e. 97.7s), the scale $\sigma = 0.19 \pm 0.01$,  and the form factor $\xi = 0.17 \pm 0.05$. Using this probability we get a prediction of 21 events above 400s, while 19 are observed. However, the same law gives the probability to get a point above $10^4$s to be $2 \times 10^{-3}$. 

We tested the addition of GRB 130925A by adding its claimed duration of 20ks \cite{pir14} obtaining a very similar result.

In conclusion these results indicate clearly that the presence of one (or two) GRB above 10ksec is an outlier.

\section{Discussion}

In the previous section we show that the light curve of ULGRBs is markedly different from that of lGRBs. Not only they display {\bf burst} activity for very long times, but also there is a "regular" afterglow decay after times larger than 10ksec (see. e.g. the statistical analysis performed in \cite{pir14}). 

Another set of evidence come from the case of GRB 130925A that showed constant radio emission over four months, the clear sign of a wind environment. Finally, as already noted, all ULGRBs display a thermal component. As an example, GRB 130925A has been observed up to 10$^7$s by XMM-Newton, with a 1keV thermal spectrum, a temperature slowly decreasing, which led \cite{pir14} to conclude that it could be the sign of a hot cocoon.

The case of GRB 130925A rises the question of whether or not we should have considered late flares in the burst duration. Late flares can be the signature of a sustained activity of the central engine. On the other hand, they can be the signs of the injection of slow blobs in the fireball that appear at later times. In any case, whatever the conclusion, the activity of the central engine for more than 20ksec needs an explanation.

Another question that arises is that of the progenitor: should we invoke a different source of energy? Our work show that ULGRBs are not an extension of normal lGRBS, and that they display also a thermail component. Accretion from an extended source seems a natural explanation, as already noted \cite{gen13, Woo12}.

\section{Conclusions}

\begin{figure*}
\begin{center}
\includegraphics[]{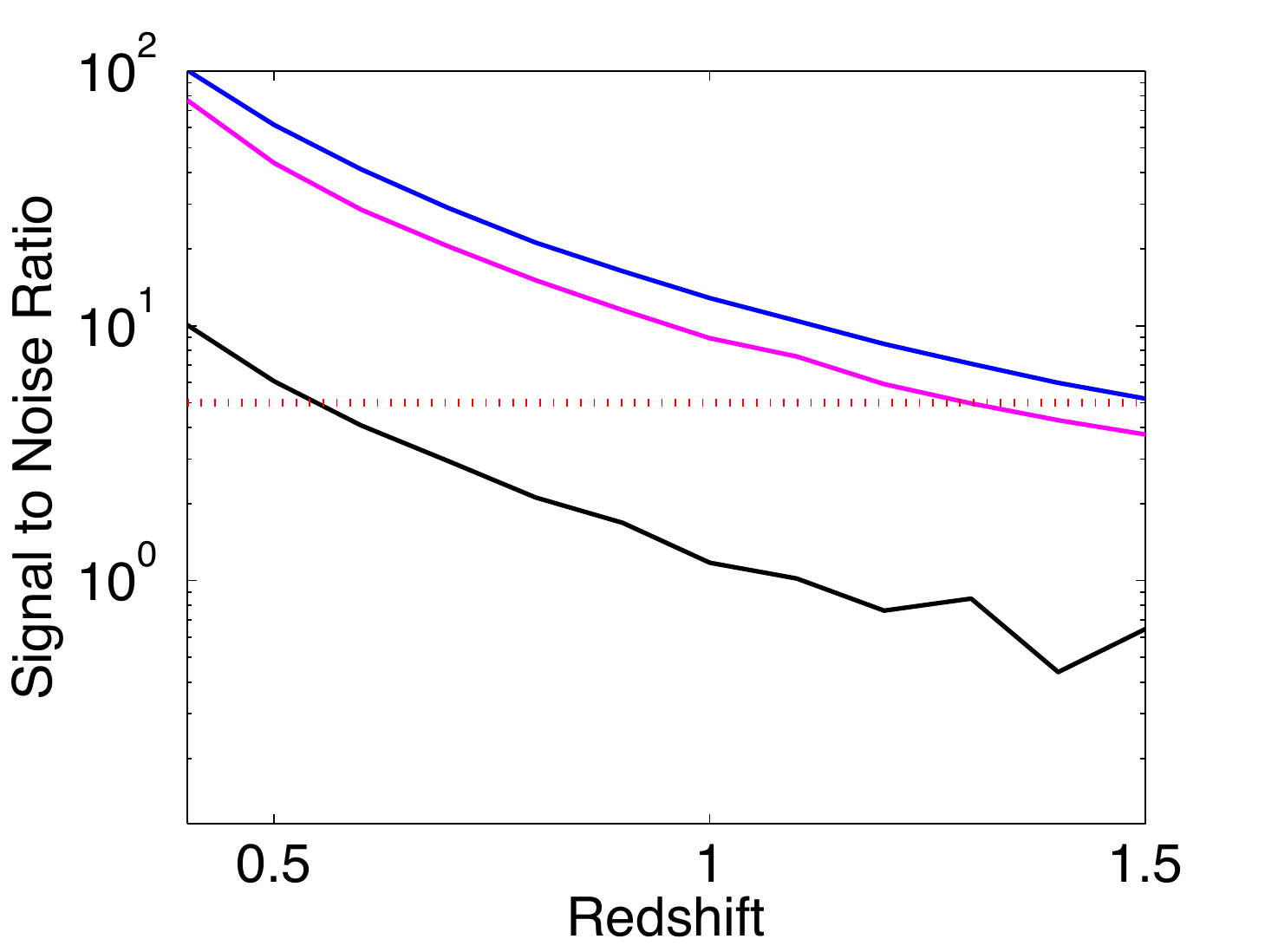} 
\caption{Detection significance expressed in signal to noise ratio units for a GRB 111209A like event as a function of its redshift. We present the detection regime of such bursts using light curve bin sizes of 1 minute (purple line) and 2 minutes (blue line); and a burst 10 times fainter with 2 minutes bin size (black line). The dotted red line indicates our detection threshold (5 sigma level), showing that even in the optimal case the maximum redshift for detection is z = 1.4-1.5. From \cite{gen13}.\label{figsens}}
\end{center}
\end{figure*}

We have presented evidences that ULGRBs form a class different of that of "regular" lGRBs. Accretion from an extended source is a natural explanation, as already noted, though probably not the only one. It is tempting to see ULGRBs as large, extended very low metallicity stars, fossils of PopIII stars, e.g. blue supergiants. 

The fact that the detected ULGRBs are all relatively nearby is not really suprising, as they can't be detected at redshift larger than 1.5 in the very best case as shown in Fig.\ref{figsens}. If we want to detect more of these guys, new experiments, such as the SVOM satellite \cite{Cor15}, should use new trigger criteria that maximize the probability to detect distant events. As ULGRBs might well be the swansong of PopIII supergiant stars, this can be considered as a priority.

\acknowledgments

This paper is under the auspices of the FIGARONet collaborative network supported by the Agence Nationale de la Recherche, program ANR-14-CE33. This work made use of data supplied by the UK Swift Science Data Centre at the University of Leicester.

\end{document}